\documentclass[conference]{IEEEtran}

\usepackage{cite}
\usepackage{amsmath,amssymb,amsfonts}
\usepackage{booktabs}
\usepackage{graphicx}
\usepackage{array}
\usepackage{textcomp}
\usepackage{xcolor}
\usepackage[hidelinks]{hyperref}

\def\BibTeX{{\rm B\kern-.05em{\sc i\kern-.025em b}\kern-.08em
    T\kern-.1667em\lower.7ex\hbox{E}\kern-.125emX}}

\begin{document}

\title{Rethinking Domain Specialization for Open-Ended Scientific Reasoning in Astronomy Language Models}

\author{
\IEEEauthorblockN{
\parbox{\textwidth}{\centering
Vanessa Lama\IEEEauthorrefmark{1},
Sanjay Das\IEEEauthorrefmark{1},
Emily Herron\IEEEauthorrefmark{1},
Yuan-Sen Ting\IEEEauthorrefmark{2}\IEEEauthorrefmark{4},
Tijmen de Haan\IEEEauthorrefmark{3},
Junqi Yin\IEEEauthorrefmark{1},
Tirthankar Ghosal\IEEEauthorrefmark{1},
Feiyi Wang\IEEEauthorrefmark{1}
}}
\IEEEauthorblockA{
\IEEEauthorrefmark{1}Oak Ridge National Laboratory, Oak Ridge, Tennessee, USA\\
\IEEEauthorrefmark{2}The Ohio State University, Columbus, Ohio, USA\\
\IEEEauthorrefmark{3}QUP/IPNS, High Energy Accelerator Research Organization (KEK),
Tsukuba, Ibaraki, Japan\\
\IEEEauthorrefmark{4}Max-Planck-Institut f\"ur Astronomie, K\"onigstuhl 17, D-69117 Heidelberg, Germany\\
Email: \{lamav,dass3,ghosalt\}@ornl.gov
}
}

\maketitle

\begin{abstract}
Domain-specialized language models are widely used for scientific question answering, but stronger general-purpose systems raise a sharper question: when does domain-specific fine-tuning remain valuable for open-ended scientific reasoning? We study this in astronomy with a curated QA benchmark from publicly available 2017--2026 Olympiad-style materials. The free-response subset contains 300 questions, including 204 text-only and 96 image-linked examples. We compare open-weight and API-served general-purpose, multimodal, and astronomy-specialized models using judge-based correctness and complementary reference metrics. Strong general-purpose models establish the highest correctness baseline in this testbed, while analyses of metric agreement, judge sensitivity, benchmark composition, and modality reveal variation not captured
by a single leaderboard. These results motivate treating domain specialization
as a task- and deployment-dependent property and highlight the role of domain-specific evaluation in determining which models, capabilities, and evaluation criteria are appropriate for scientific workflows.
\end{abstract}


\section{Introduction}

Large language models (LLMs) and multimodal foundation models are increasingly used for scientific question answering, literature understanding, and knowledge-intensive reasoning. Astronomy provides a concrete testbed for broader AI-for-science evaluation because its problems combine specialized terminology, quantitative reasoning, diagrams, and physical assumptions. Rigorous evaluation remains difficult because open-ended problems are poorly served by exact match or shallow overlap metrics.

A central question for scientific language modeling is not whether domain specialization is useful, but when it provides measurable advantages over strong general-purpose models. Astronomy-specialized models have improved over earlier general baselines, while open or locally served models raise deployment questions not captured by accuracy alone. At the same time, general-purpose models have advanced rapidly on mathematical, scientific, expert-level, and olympiad-style reasoning benchmarks~\cite{rein2023gpqa,wang2024scibench,yue2024mmmu,he2024olympiadbench}. Domain fine-tuning should therefore be compared with current general-purpose systems, not only smaller or older baselines. Most astronomy-focused evaluations emphasize multiple-choice QA or similar datasets, where mapping response to a definitive solution can provide context on model performance. However, we have yet to fully capture free-response reasoning with explanations, intermediate steps, and quantitative or conceptual conclusions.

To examine this gap, we study a curated astronomy QA benchmark derived from Olympiad-style materials spanning 2017--2026. We focus on the free-response subset, including text-only and image-linked questions, while a multiple-choice track is reserved for future expansion of our study. Our evaluation complements broader scientific and multimodal benchmarks such as MMLU, ScienceQA, GPQA, SciBench, MMMU, and OlympiadBench~\cite{hendrycks2021mmlu,lu2022scienceqa,rein2023gpqa,wang2024scibench,yue2024mmmu,he2024olympiadbench} by focusing on open-ended astronomy problems with reference solutions, multi-part reasoning structure, and image-linked scientific context. We evaluate open-weight and API-served models with general-purpose, multimodal, reasoning, and astronomy-specialized capabilities. 

We use this benchmark to characterize scientific reasoning across model types, question properties, modalities, and evaluation methods.

The contributions of this paper are:
\begin{itemize}
  \item a structured 2017--2026 free-response astronomy QA corpus and conversion
  pipeline preserving subpart context, reference solutions, and associated figures;

  \item a unified benchmarking framework for comparing general-purpose,
  multimodal, and astronomy-specialized language models on open-ended
  astrophysics reasoning tasks;

  \item a systematic characterization of model behavior across question
  properties, evaluation metrics, judge selection, modality, and benchmark
  robustness.
\end{itemize}

\begin{figure*}[!t]
  \centering
  \includegraphics[width=0.85\textwidth]{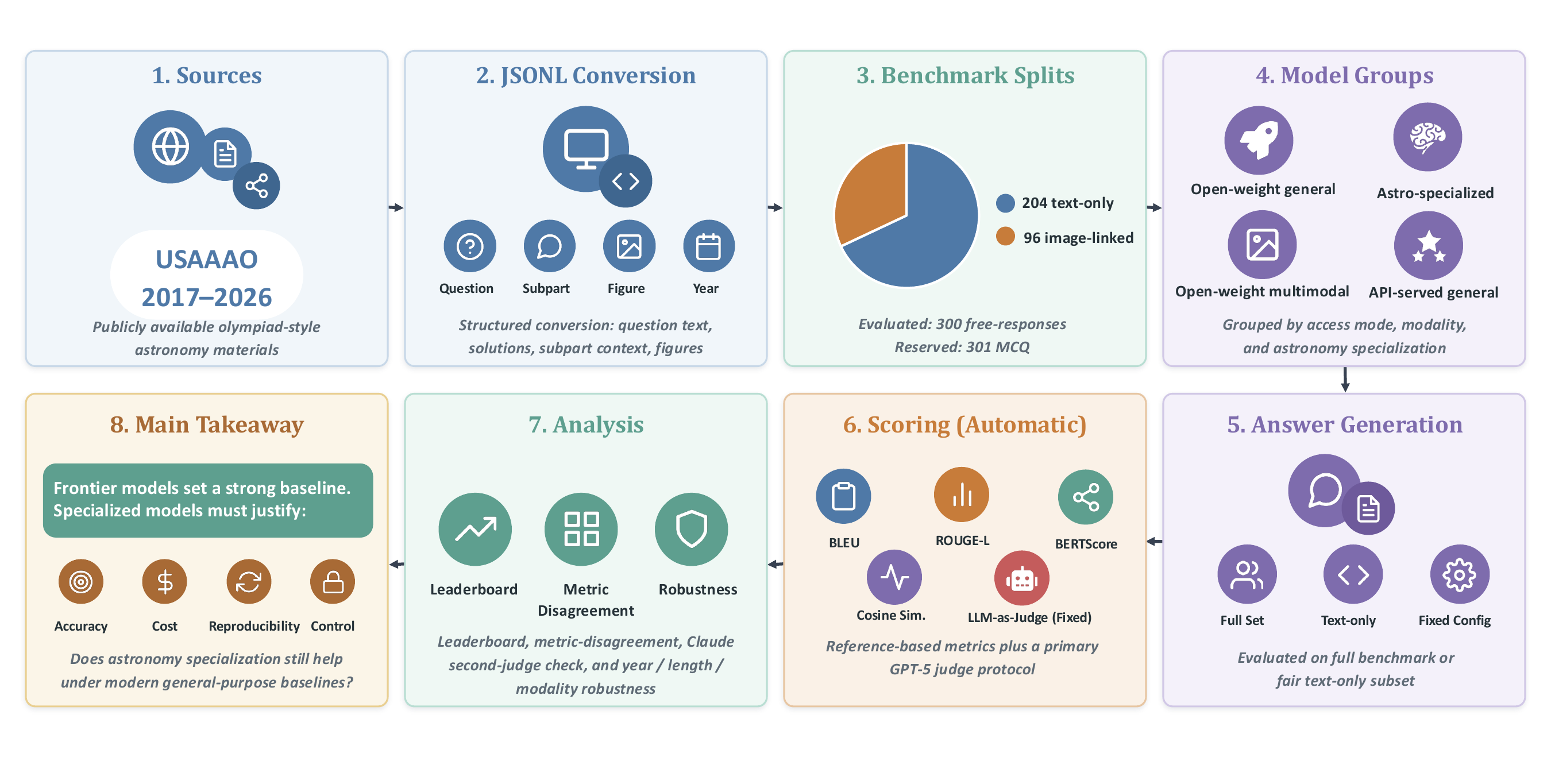}
  \caption{Experimental pipeline for constructing the benchmark, generating model
  responses, evaluating scientific answer quality, and analyzing model behavior
  across performance, metric-agreement, robustness, and modality views.}
  \label{fig:experiment-pipeline}
\end{figure*}

Fig.~\ref{fig:experiment-pipeline} summarizes the experimental pipeline from
benchmark construction through model response generation, evaluation, and
exploratory analysis.

\section{Background and Related Work}
Scientific question answering has evolved from broad knowledge-based benchmarks toward increasingly specialized evaluations of domain knowledge, complex reasoning, and multimodal understanding. Yet evaluating language models remains a moving target as models, data, capabilities, and scientific use cases rapidly evolve. As AI becomes more integrated into scientific practice, evaluation must likewise mature toward systematic frameworks that identify which measures of model capability and reliability are most meaningful for a given scientific workflow.

\subsection{Scientific Question Answering and Benchmarking}
Scientific QA benchmarks test reasoning over technical material rather than surface retrieval. MMLU, ScienceQA, GPQA, and SciBench cover broad academic knowledge, multimodal science reasoning, expert questions, and college-level problem solving~\cite{hendrycks2021mmlu,lu2022scienceqa,rein2023gpqa,wang2024scibench}. MMMU and OlympiadBench emphasize expert-level visual and olympiad-style reasoning~\cite{yue2024mmmu,he2024olympiadbench}. Particularly in astronomy, AstroMLab 1 introduced a multiple-choice benchmark~\cite{ting2025astromlab1}. As an extension, our benchmark focuses on open-ended problems requiring physical modeling, derivation, numerical estimation, and explanation in an attempt to evaluate a separate aspect of scientific QA.

\subsection{Evaluation of Open-Ended Model Outputs}
Open-ended scientific responses are difficult to evaluate with exact match or multiple-choice accuracy. BLEU and ROUGE-L provide inexpensive and reproducible measures of response similarity, but primarily capture lexical overlap rather than scientific correctness~\cite{papineni2002bleu,lin2004rouge}. BERTScore and sentence-embedding cosine similarity capture semantic similarity beyond lexical overlap, but may still assign high similarity to scientifically incorrect responses~\cite{zhang2020bertscore,reimers2019sentencebert}. LLM-as-judge evaluation provides a practical approach for assessing open-ended responses, but its reliability depends on factors such as judge selection, evaluation criteria, and potential biases~\cite{zheng2023llmjudge,liu2023geval,wataoka2024selfpreference}. SciTrust and EAIRA similarly argue for multi-faceted scientific evaluation~\cite{herron2024scitrust,cappello2025eaira}. We therefore complement judge-based evaluation with reference metrics and robustness analyses to provide a more comprehensive assessment of model performance.

\subsection{Domain-Specialized and Multimodal Models for Science}
Domain-adapted language models seek to improve scientific capabilities through continued pretraining or instruction tuning on specialized corpora. In astronomy, AstroLLaMA introduced models trained on astronomy literature~\cite{nguyen2023astrollama}, while subsequent AstroMLab efforts expanded domain-specific benchmarking and model development through the AstroLLaMA and AstroSage families~\cite{pan2024astromlab2,dehaan2024astromlab3,dehaan2025astromlab4,dehaan2025astrosage}. AstroSage further represents domain adaptation at scale, with its 70B-scale model trained using leadership-class computing resources on Frontier~\cite{dehaan2025astrosage}. These efforts motivate evaluating specialization not as an assumed advantage, but alongside general-purpose models and practical considerations such as open-weight access, local deployment, and data governance. We also consider multimodal capability, as astronomy problems frequently incorporate diagrams, plots, and sky maps.

\section{Methodology}

\subsection{Dataset Source and Construction}
The dataset is derived from publicly available USA Astronomy and Astrophysics Organization (USAAAO) Olympiad materials spanning 2017--2026. Annual \LaTeX{} question and solution files are processed through a structured conversion pipeline to produce JSONL records that preserve question text, reference solutions, subpart organization, year identifiers, and associated figures. The resulting free-response dataset contains 300 examples.

\subsection{Record Construction}
Each JSONL record represents a standalone problem or individually evaluated subpart and includes the year, problem identifier, question text, reference answer, length category, and associated image paths when applicable. The core schema is \texttt{\{year, id, question, reference\_answer, image\_paths, length\_bucket\}}, with additional metadata used to preserve parent-problem relationships, subpart structure, source information, and figures. Length categories are assigned heuristically based on problem structure, number of sub-parts, and expected reasoning depth rather than fixed token thresholds, and are used only for robustness analysis. For multi-part problems, shared context is retained to preserve the interpretation of equations, constants, diagrams, and coordinate conventions. Representative records are provided in the Appendix.

\subsection{Evaluation Splits}
The curated free-response dataset contains 204 text-only and 96 image-linked examples, with their distribution across years shown in Fig.~\ref{fig:data-overview}. We define two evaluation settings based on model modality. The \emph{text-only} setting includes the 204 examples that do not require visual input, providing a common evaluation set for all models. The
\emph{full} setting includes all 300 examples and is used to evaluate models capable of processing both textual and visual inputs. This separation enables comparisons across model classes without disadvantaging text-only models while retaining the multimodal characteristics of the benchmark. First-round multiple-choice materials are reserved for future development as a
complementary accuracy-based evaluation track.

\section{Experimental Setup}

\subsection{Dataset}
\begin{figure*}[!t]
  \centering
  \includegraphics[width=\textwidth]{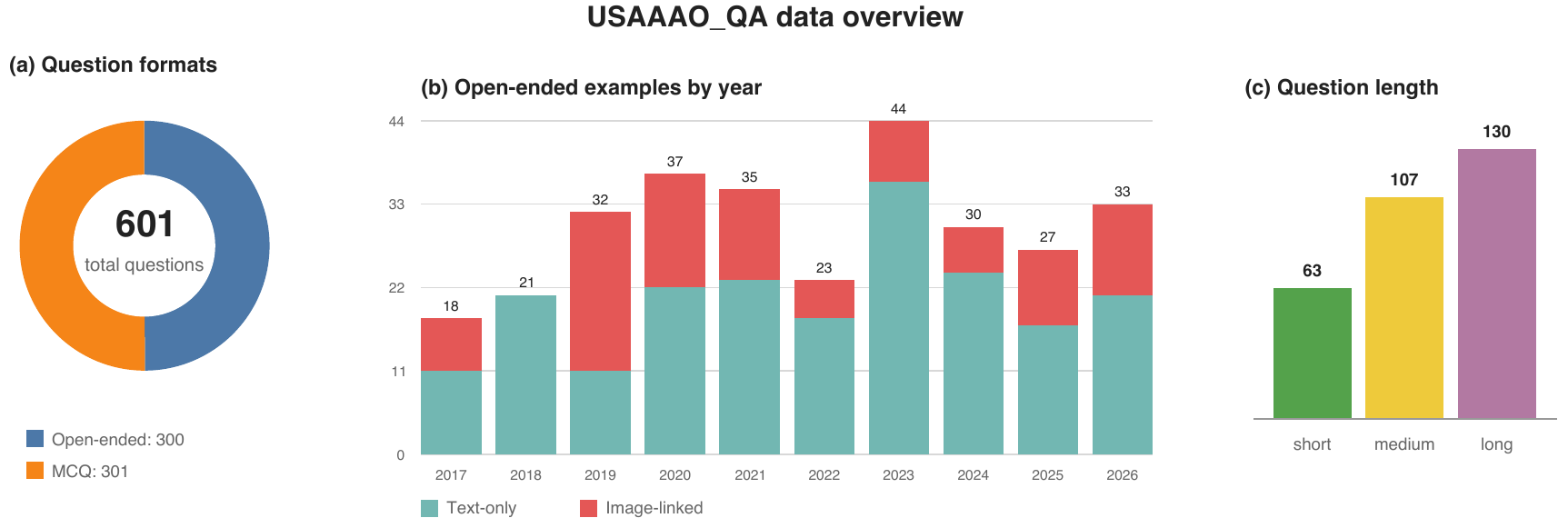}
  \caption{Overview of the USAAAO\_QA corpus. (a) Distribution of the 601 questions between free-response and multiple-choice formats. (b) Distribution of the 300 free-response examples across years and text-only or image-linked modalities. (c) Distribution of the free-response examples across short, medium, and long length categories. The 301 multiple-choice questions are reserved for a future benchmark track.}
  \label{fig:data-overview}
\end{figure*}

Fig.~\ref{fig:data-overview} summarizes the overall dataset composition, including question format, year and modality distributions, and question-length categories.

\subsection{Models}
The evaluated systems span general-purpose, multimodal, and
astronomy-specialized language models under both local and API-served
configurations (Fig.~\ref{fig:experiment-pipeline}). Open-weight models include
Llama and GPT-OSS general-purpose models, Gemma multimodal models, and the
astronomy-adapted AstroLLaMA and AstroSage families. API-served models include
GPT-5.5, Claude Sonnet 4.6, and GPT-OSS 120B.

Evaluation follows model input capability and serving support. Text-only models
use the shared 204-example subset, while models supporting image inputs use the
full 300-example benchmark.

\subsection{Inference Configuration}
All models use a common generation and evaluation pipeline where supported. Each response is generated once and evaluated against its reference solution using automatic metrics and a fixed GPT-5 judge, which assigns scores of correct (1.0), partially correct (0.5), or incorrect (0.0). GPT-5.5 is evaluated without additional reasoning effort, while GPT-OSS 120B uses the provider-required low-reasoning setting. Higher-reasoning configurations are reserved for future ablation studies.

\subsection{Computational Platform}
Local open-weight inference was performed on an 8-GPU NVIDIA A100 system using a CUDA-enabled PyTorch environment. The benchmark is inference-only; no training or fine-tuning was performed as part of this study. Models were allocated one or multiple GPUs according to memory requirements, with independent evaluations executed concurrently when resources permitted. Per-example predictions, metrics, judge decisions, and run metadata were retained, with resumable execution used to limit repeated inference following interruptions.

API-served models were accessed through Oak Ridge National Laboratory (ORNL) Azure OpenAI and Oak Ridge Leadership Computing Facility (OLCF)/American Science Cloud (AmSC) OpenAI-compatible endpoints. Model identifiers, access dates, and run metadata were retained for reproducibility, while data loading, metric computation, serialization, and judge evaluation were handled through the same benchmark pipeline used for local models. Provider-side GPU utilization and energy measurements were not available; local evaluations retained wall-clock and per-example execution records.

\begin{table}[!t]
  \caption{Compute and serving configurations used for benchmark inference.}
  \label{tab:compute-config}
  \centering
  \begin{tabular}{p{0.28\columnwidth}p{0.20\columnwidth}p{0.40\columnwidth}}
    \toprule
    Model group & Serving & Compute configuration \\
    \midrule
    Open 1B--20B text models
      & Local
      & A100 80GB; typically single-GPU inference \\
    Open 70B text models
      & Local
      & A100 80GB; multi-GPU inference as required \\
    Open multimodal models
      & Local
      & A100 80GB; image-enabled inference on the full benchmark \\
    GPT-5 judge and GPT-5.5
      & API
      & ORNL Azure OpenAI; provider-managed compute \\
    Claude Sonnet 4.6 and GPT-OSS 120B
      & API
      & OLCF/AmSC OpenAI-compatible endpoint; provider-managed compute \\
    \bottomrule
  \end{tabular}
\end{table}

\subsection{Evaluation Protocol and Metrics}
Models are evaluated on the full or text-only benchmark according to their input capabilities, with comparisons made primarily within the same setting. Free-response answers are assessed using a primary LLM-as-judge score alongside complementary reference-based metrics. Let $y$ denote the reference answer and $\hat{y}$ the generated response. Then,

\paragraph{Judge score.}
The primary correctness metric is a fixed GPT-5 judge score,
\[
s_{\mathrm{judge}}(\hat{y},y)\in\{0,0.5,1\},
\]
corresponding to incorrect, partially correct, and correct responses,
respectively. We report the mean judge score
\[
\bar{s}_{\mathrm{judge}}
=\frac{1}{N}\sum_{i=1}^{N}s_{\mathrm{judge}}(\hat{y}_i,y_i).
\]

\paragraph{Reference metrics.}
We complement judge-based correctness with BLEU, ROUGE-L, BERTScore, and
sentence-embedding cosine similarity. BLEU measures modified $n$-gram precision
with a brevity penalty,
\[
\mathrm{BLEU}
=\mathrm{BP}\exp\left(\sum_{n=1}^{4}w_n\log p_n\right),
\]
where $p_n$ denotes modified $n$-gram precision and $w_n$ are uniform weights.
ROUGE-L measures longest-common-subsequence overlap between the generated and
reference responses. BERTScore instead compares contextual token embeddings
through similarity-based matching; we report baseline-rescaled F1 scores, which
may take negative values. Sentence-level semantic similarity is computed as
\[
\mathrm{CosSim}(\hat{y},y)=
\frac{e(\hat{y})\cdot e(y)}
{\|e(\hat{y})\|\,\|e(y)\|},
\]
where $e(\cdot)$ denotes the sentence-embedding model. We treat these
reference-based metrics as complementary diagnostic measures rather than
direct measures of scientific correctness.

\subsection{Score Uncertainty and Robustness}
Because benchmark composition varies by year, modality, and length category, we
quantify uncertainty in aggregate model scores using nonparametric bootstrap
resampling. For each replicate $b$, we sample $N$ examples with replacement and
compute
\[
\bar{s}^{(b)}_{\mathrm{judge}}
= \frac{1}{N}\sum_{i=1}^{N}s_i^{(b)}.
\]
Confidence intervals are obtained from the empirical distribution of the
bootstrap means.

We further examine robustness across benchmark characteristics. Year-level
variation is summarized as
\[
\sigma^2_{\mathrm{year}}
= \frac{1}{|\mathcal{Y}|}\sum_{y\in\mathcal{Y}}
(\bar{s}_y-\bar{s})^2,
\]
where $\bar{s}_y$ and $\bar{s}$ denote the year-specific and overall mean judge
scores, respectively. For models evaluated on the full benchmark, the modality
gap is
\[
\Delta_{\mathrm{modality}}
= \bar{s}_{\mathrm{text}}-\bar{s}_{\mathrm{image}},
\]
where positive values indicate lower performance on image-linked questions.
Together, these analyses assess the stability of aggregate model comparisons
across benchmark composition. Bootstrap intervals capture variation from example
resampling only, as each model generates one response per question.

\section{Results}

We evaluate domain specialization relative to strong general-purpose baselines
through complementary views of aggregate performance, modality, metric agreement,
judge sensitivity, and robustness.

\subsection{Aggregate Model Performance}

\begin{table*}[!t]
  \caption{Free-response benchmark results for 2017--2026. \emph{Judge} reports
  performance on each model's evaluation setting, while
  \emph{Judge$_{\mathrm{text}}$} reports performance on the shared 204-example
  text-only subset. Bold values indicate the highest full-benchmark and shared
  text-only judge scores.}
  \label{tab:usaaao-results}
  \centering
  \footnotesize
  \setlength{\tabcolsep}{3pt}
  \begin{tabular}{llccccccc}
    \toprule
    Model & Setting & $N$ & Judge & Judge$_{\mathrm{text}}$ & BLEU &
    ROUGE-L & BERTScore F1 & Cosine Sim. \\
    \midrule
    Gemma 3 4B IT            & Full      & 300 & 0.0967 & 0.0931 & 0.0143 & 0.1757 & -0.0336 & 0.5860 \\
    Gemma 4 26B A4B IT       & Full      & 300 & 0.3567 & 0.3897 & 0.0161 & 0.1828 & -0.0311 & 0.5980 \\
    Gemma 4 31B IT           & Full      & 300 & 0.3933 & 0.4314 & 0.0155 & 0.1872 & -0.0237 & 0.6064 \\
    \textbf{GPT-5.5}         & Full      & 300 & \textbf{0.6700} & \textbf{0.7083} & 0.0127 & 0.2168 & -0.0592 & 0.6072 \\
    Claude Sonnet 4.6        & Text-only & 204 & 0.6814 & 0.6814 & 0.0115 & 0.1908 & -0.0649 & 0.6077 \\
    GPT-OSS 120B             & Text-only & 204 & 0.6544 & 0.6544 & 0.0120 & 0.1899 & -0.1162 & 0.5910 \\
    Llama 3.2 1B Instruct    & Text-only & 204 & 0.0343 & 0.0343 & 0.0175 & 0.1724 & -0.0447 & 0.5725 \\
    Llama 3.1 8B             & Text-only & 204 & 0.0172 & 0.0172 & 0.0187 & 0.1312 & -0.0800 & 0.5162 \\
    Llama 3 8B Instruct      & Text-only & 204 & 0.1201 & 0.1201 & 0.0209 & 0.1758 & -0.0736 & 0.5894 \\
    Llama 3.1 70B Instruct   & Text-only & 204 & 0.2598 & 0.2598 & 0.0224 & 0.2058 & -0.0023 & 0.6193 \\
    GPT-OSS 20B              & Text-only & 204 & 0.3284 & 0.3284 & 0.0114 & 0.1478 & -0.1476 & 0.5988 \\
    AstroLLaMA 3 8B AIC      & Text-only & 204 & 0.0833 & 0.0833 & 0.0208 & 0.1403 & -0.0607 & 0.5220 \\
    AstroSage 8B             & Text-only & 204 & 0.1005 & 0.1005 & 0.0167 & 0.1280 & -0.0680 & 0.5307 \\
    AstroSage 70B (20251009) & Text-only & 204 & 0.3333 & 0.3333 & 0.0204 & 0.1966 &  0.0034 & 0.5997 \\
    \bottomrule
  \end{tabular}
\end{table*}

Table~\ref{tab:usaaao-results} gives aggregate performance across the 2017--2026 free-response benchmark. API-served general-purpose models set the highest baseline: GPT-5.5 reaches 0.6700 on the full benchmark, while Claude Sonnet 4.6 and GPT-OSS 120B reach 0.6814 and 0.6544 on the text-only subset. Among open-weight multimodal models, Gemma 4 31B IT is strongest at 0.3933. Because this view mixes input settings, we next isolate the shared text-only subset.

\subsection{Text-Only Comparison}

To control for modality, we compare all models on the shared 204-example text-only
subset. GPT-5.5, Claude Sonnet 4.6, and GPT-OSS 120B achieve judge scores of
0.7083, 0.6814, and 0.6544, respectively. Among locally evaluated open-weight
models, AstroSage 70B and GPT-OSS 20B obtain similar scores of 0.3333 and 0.3284.
This comparison provides a common setting for examining the effect of domain
specialization relative to general-purpose baselines. Appendix
Table~\ref{tab:specialization-deltas} further reports specialization deltas against
comparable general models. We next examine whether reference-based metrics reflect the same performance patterns observed under judge-based evaluation.

\subsection{Metric Disagreement}
\begin{figure}[!t]
  \centering
  \includegraphics[width=\linewidth]{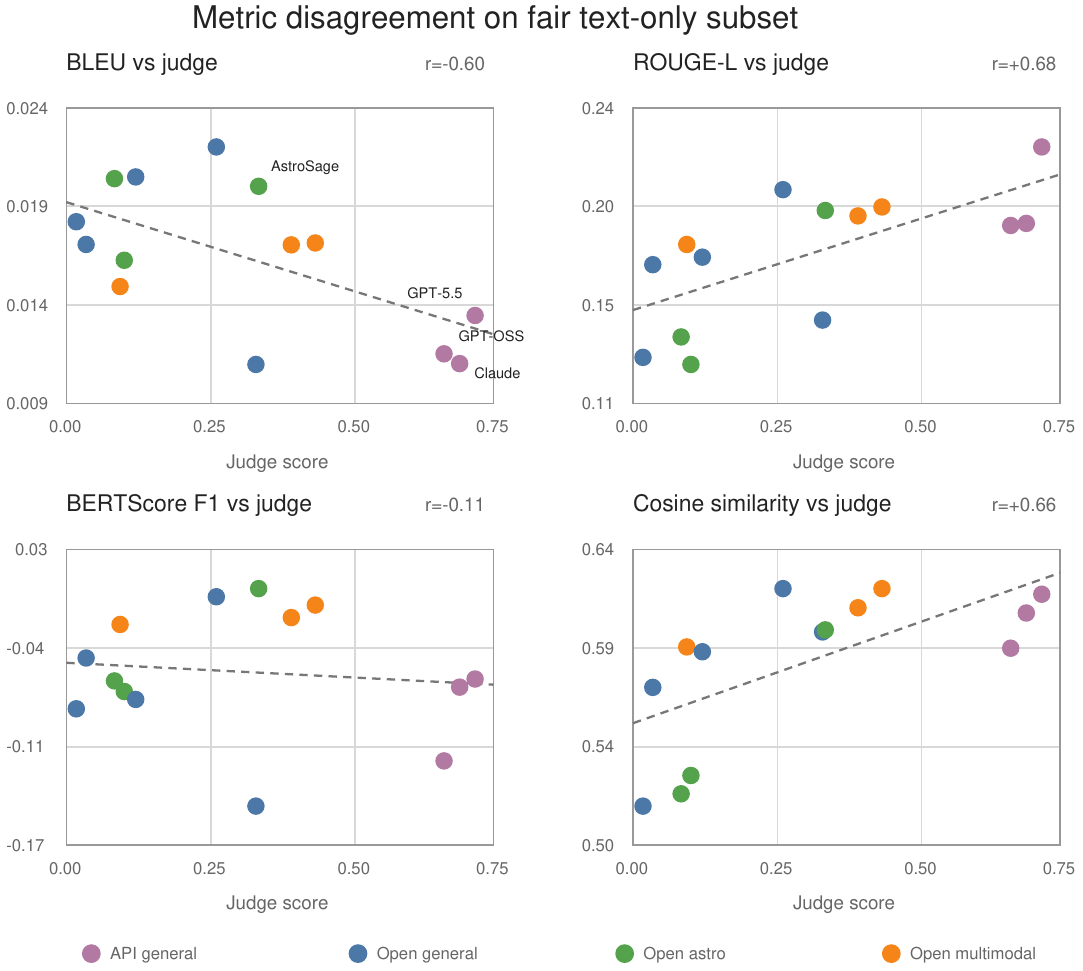}
  \caption{Relationship between judge scores and reference-based metrics on the
  shared text-only subset. Correlations vary considerably across BLEU, ROUGE-L,
  BERTScore F1, and sentence-level cosine similarity, illustrating that
  reference similarity does not consistently track judged correctness.}
  \label{fig:metric-disagreement}
\end{figure}

Fig.~\ref{fig:metric-disagreement} shows
substantial variation across metrics on the shared text-only subset. ROUGE-L and
sentence-level cosine similarity show moderate positive correlations with judge
score ($r=0.68$ and $r=0.66$, respectively), whereas BERTScore shows little
association ($r=-0.11$). BLEU is negatively correlated with judge score
($r=-0.60$), with scores remaining low even among models receiving substantially
higher correctness judgments.

These differences are particularly relevant when comparing domain-specialized
and general-purpose models. Domain-specific terminology may increase lexical or
semantic similarity to reference solutions without necessarily indicating a
correct solution, while correct responses may follow alternative wording or
derivations. We therefore treat reference-based metrics as complementary
diagnostics and use judge-based scoring as the primary correctness measure. Given these differences across metrics, we assess the sensitivity of the evaluation to judge selection.

\subsection{Judge Sensitivity}

To assess whether model comparisons depend on evaluator choice, we re-score the five highest-performing eligible models under the primary GPT-5 judge on the shared 204-example text-only benchmark using Claude Sonnet 4.6 as a second judge. Claude Sonnet 4.6 is excluded to avoid self-evaluation.

\begin{table}[!t]
  \caption{Second-judge sensitivity on the shared text-only subset. $\Delta$
  denotes the Claude score minus the GPT-5 score; agreement is the fraction of
  examples receiving identical 0/0.5/1 scores.}
  \label{tab:judge-sensitivity}
  \centering
  \small
  \setlength{\tabcolsep}{3pt}
  \begin{tabular}{lccccc}
    \toprule
    Model & $N$ & GPT-5 & Claude & $\Delta$ & Agree \\
    \midrule
    GPT-5.5        & 204 & 0.7083 & 0.6936 & -0.0147 & 0.868 \\
    GPT-OSS 120B   & 204 & 0.6544 & 0.6691 & +0.0147 & 0.882 \\
    Gemma 4 31B IT & 204 & 0.4314 & 0.5319 & +0.1005 & 0.623 \\
    AstroSage 70B  & 204 & 0.3333 & 0.3627 & +0.0294 & 0.868 \\
    GPT-OSS 20B    & 204 & 0.3284 & 0.4877 & +0.1593 & 0.608 \\
    \bottomrule
  \end{tabular}
\end{table}

Table~\ref{tab:judge-sensitivity} shows that judge selection affects score
magnitude more strongly for some models than others, while the highest-performing
models remain stable across both evaluators. Agreement ranges from 0.608 for
GPT-OSS 20B to 0.882 for GPT-OSS 120B, with score differences ranging from
$-0.0147$ to $+0.1593$. Because the primary GPT-5 judge shares a model family
with some evaluated systems, same-family evaluator preference is a potential
confound~\cite{wataoka2024selfpreference}. However, the observed score shifts do
not indicate a uniform advantage for models from the same family. These results
reinforce the need to interpret small score differences cautiously and motivate
examining whether model comparisons remain stable across variations in benchmark
composition.

\subsection{Performance Robustness}

\begin{figure}[!t]
  \centering
  \includegraphics[width=\linewidth]{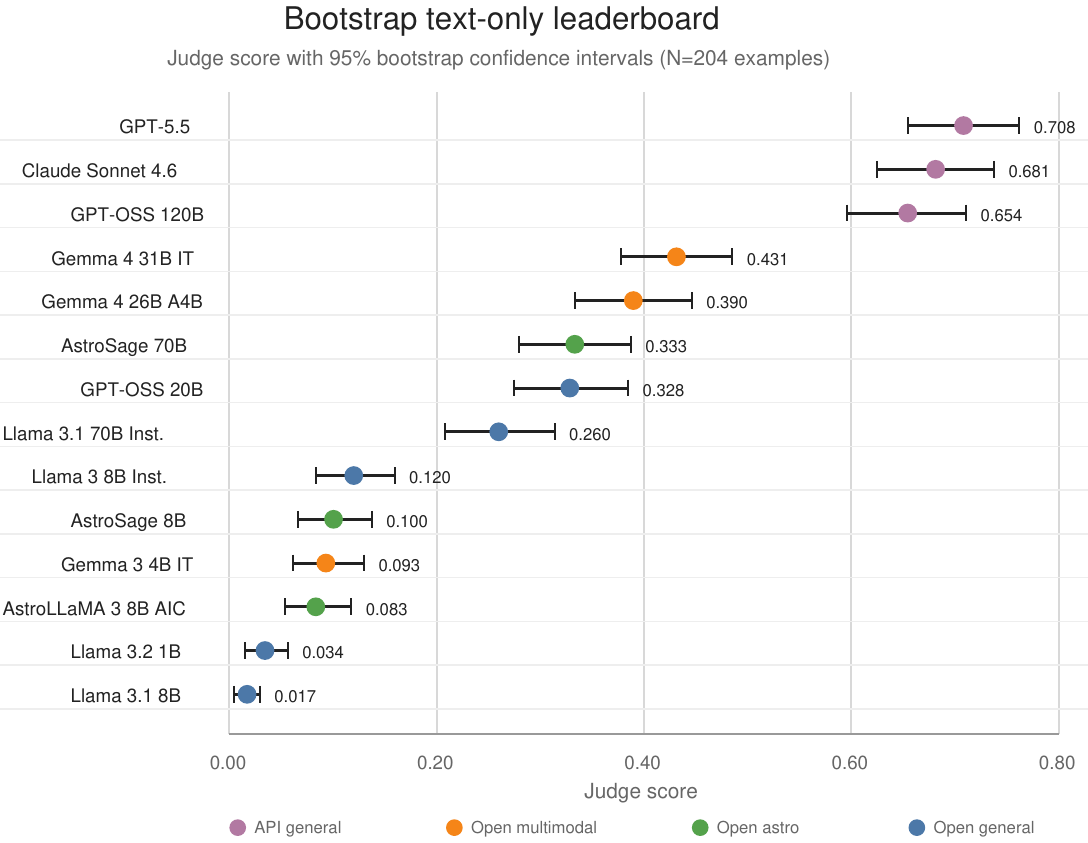}
  \caption{Bootstrap estimates of model performance on the shared text-only
  subset ($N=204$). Points denote mean judge scores and intervals show 95\%
  nonparametric bootstrap confidence intervals.}
  \label{fig:bootstrap-leaderboard}
\end{figure}

We first assess whether differences in aggregate performance exceed variation
induced by benchmark composition. Fig.~\ref{fig:bootstrap-leaderboard} shows
95\% bootstrap confidence intervals for judge scores on the shared text-only
subset. Large performance differences remain well separated under resampling,
whereas intervals overlap for closer comparisons, including GPT-5.5 and Claude
Sonnet 4.6 and AstroSage 70B and GPT-OSS 20B. These results suggest that small
differences in aggregate scores should not be interpreted as definitive model
rankings.

\begin{figure}[!t]
  \centering
  \includegraphics[width=\linewidth]{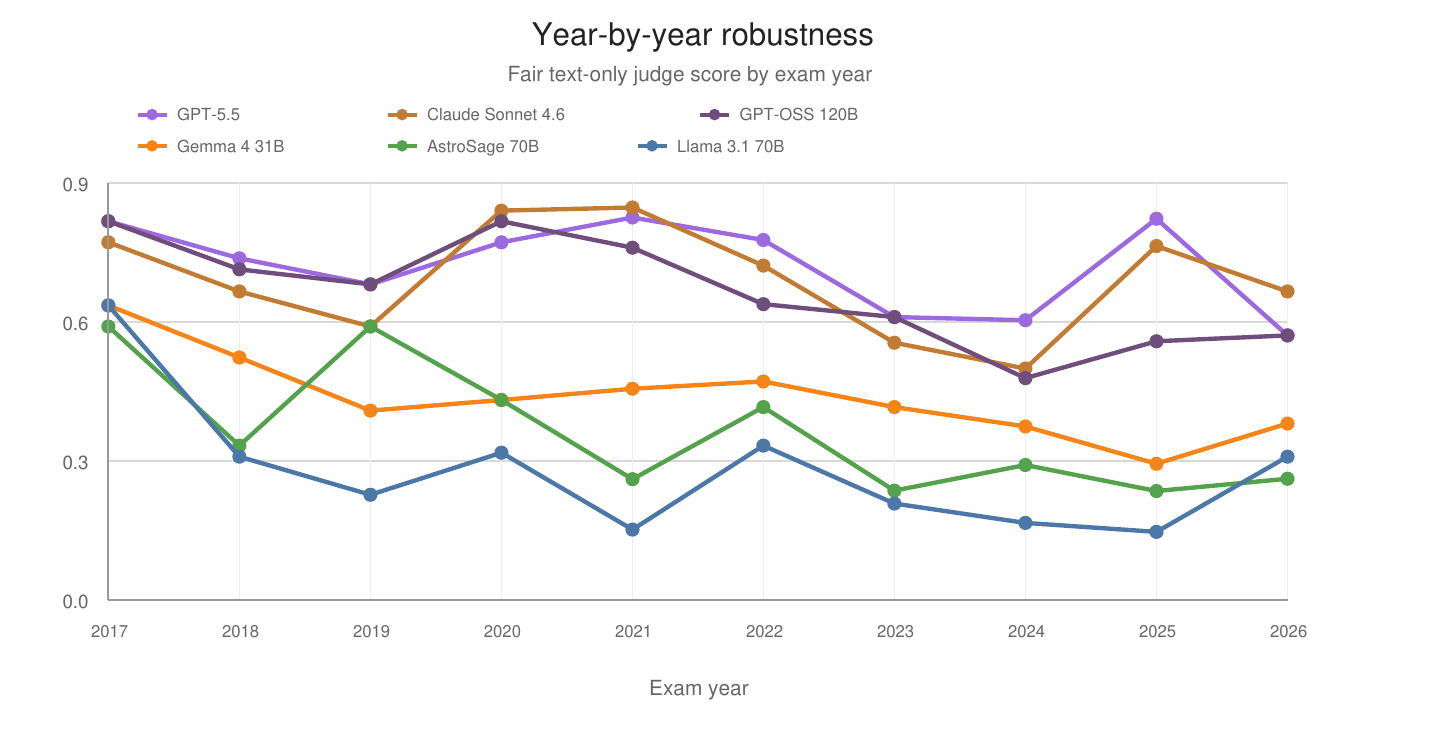}
  \caption{Year-level performance on the shared text-only subset. Lines show
  mean judge scores by exam year for selected models.}
  \label{fig:yearly-robustness}
\end{figure}

We further examine variation across exam year and question-length category.
Fig.~\ref{fig:yearly-robustness} shows that performance varies across years,
indicating sensitivity to changes in benchmark composition and problem content.
Despite this variation, broader differences among model groups are generally
preserved.

\begin{figure}[!t]
  \centering
  \includegraphics[width=\linewidth]{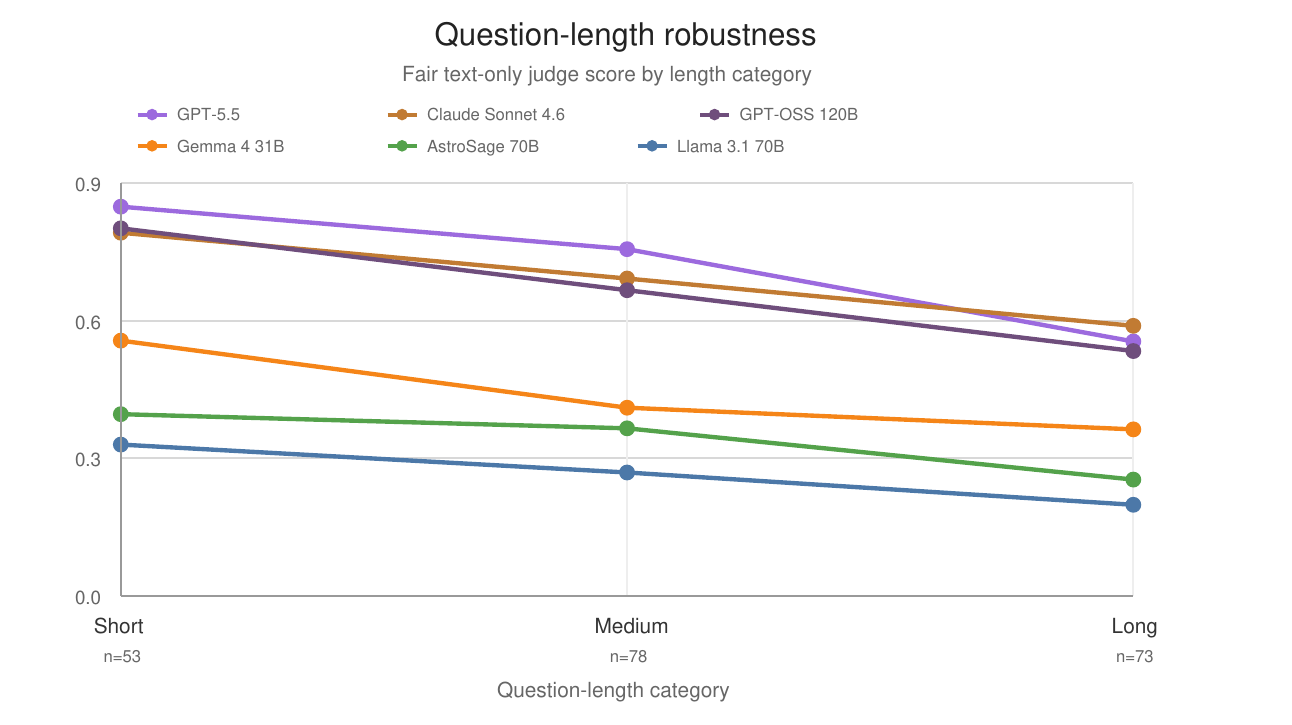}
  \caption{Performance across question-length categories on the shared text-only
  subset. Lines show mean judge scores for short, medium, and long problems
  ($N=53$, $78$, and $73$, respectively).}
  \label{fig:length-robustness}
\end{figure}

Question length produces a more consistent pattern. As shown in
Fig.~\ref{fig:length-robustness}, judge scores decline from short to medium and
long problems across all selected models. Because the length categories also
reflect problem structure and expected reasoning depth, this trend provides a
coarse indication of increasing difficulty for longer, more involved problems.

\subsection{Modality Effects}

\begin{figure}[!t]
  \centering
  \includegraphics[width=\linewidth]{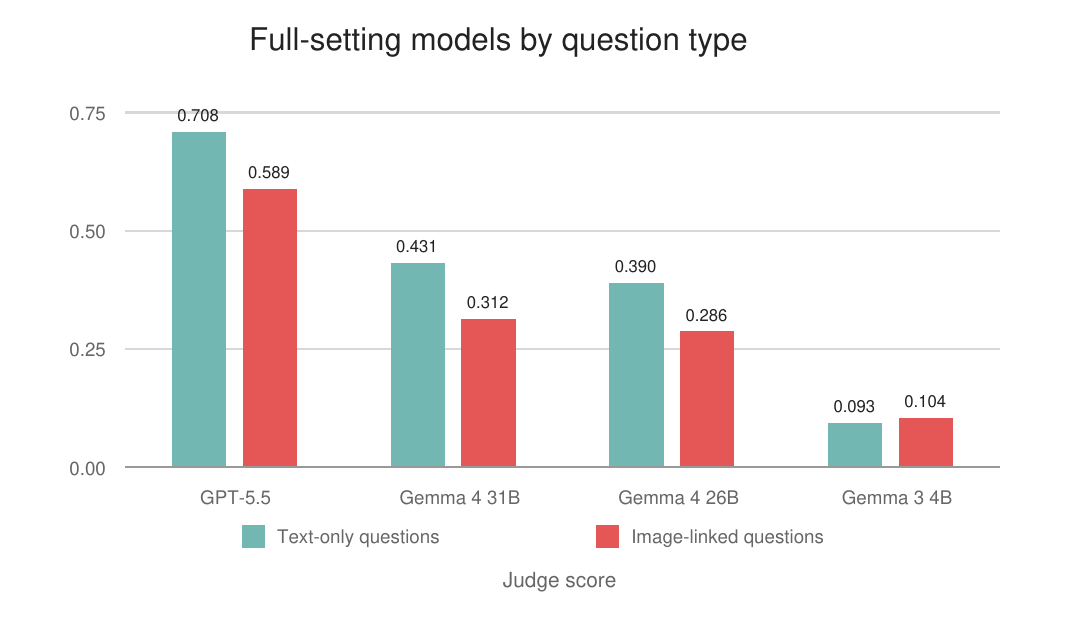}
  \caption{Performance by question modality for models evaluated on the full
  benchmark. Bars show mean judge scores on the 204 text-only and 96
  image-linked questions.}
  \label{fig:modality-breakdown}
\end{figure}

Furthermore, we examine whether performance differs between text-only and image-linked
questions for models evaluated on the full benchmark.
Fig.~\ref{fig:modality-breakdown} shows lower judge scores on image-linked
questions for GPT-5.5 and the larger Gemma variants. GPT-5.5 decreases from
0.708 on text-only questions to 0.589 on image-linked questions, while Gemma 4
31B and Gemma 4 26B decrease from 0.431 to 0.312 and from 0.390 to 0.286,
respectively. Gemma 3 4B remains substantially lower in both settings.

The consistent gap among the higher-performing full-setting models suggests that
image-linked problems introduce additional challenges beyond those observed on
the text-only subset. However, modality is confounded with problem content and
difficulty, so these differences should not be attributed solely to visual
reasoning capability. A controlled multimodal comparison across a broader set of
vision-capable models would be needed to isolate this effect.

\section{Error Analysis}

We conduct a targeted qualitative error analysis on a subset of model responses
to characterize failures not fully captured by aggregate correctness or
reference-based similarity metrics. The annotated responses reveal four recurring
categories: incomplete derivations, numerical or unit errors, visual grounding
failures, and context fragmentation. Table~\ref{tab:error-buckets} summarizes
these failure patterns.

\begin{table}[!t]
  \caption{Failure patterns identified through targeted annotation of model responses.}
  \label{tab:error-buckets}
  \centering
  \small
  \setlength{\tabcolsep}{4pt}
  \renewcommand{\arraystretch}{1.15}
  \begin{tabular}{p{0.30\columnwidth}p{0.60\columnwidth}}
    \toprule
    Failure type & Description \\
    \midrule
    Incomplete derivation
      & Correct concept or relation is identified, but the reasoning or
        calculation is not completed. \\

    Numerical / unit error
      & Reasoning is plausible, but errors occur in algebra, units, constants,
        or numerical scale. \\

    Visual grounding
      & Required information from a figure, plot, or diagram is ignored or
        incorrectly interpreted. \\

    Context fragmentation
      & A subpart is answered without necessary context or results established
        earlier in the problem. \\
    \bottomrule
  \end{tabular}
\end{table}

These failure patterns provide additional context for the metric disagreement
observed in Fig.~\ref{fig:metric-disagreement}. A response may use appropriate
domain terminology or begin from the correct physical relation while still
producing an incomplete derivation or incorrect numerical result. Similarly,
errors involving visual evidence or shared problem context may not be adequately
captured by lexical or embedding-based similarity. These observations reinforce
the use of reference-based metrics as complementary diagnostics rather than
stand-alone measures of scientific correctness.

Because the error annotation covers only a subset of responses, we treat this
analysis as qualitative rather than as an estimate of failure prevalence across
the benchmark. Extending the annotation across models and question categories
would enable a more systematic comparison of how failure modes vary with model
specialization, modality, and problem characteristics.

\section{Discussion}

Our results suggest that the value of domain specialization should be established
through task- and deployment-specific evaluation rather than assumed from domain
adaptation alone. In this astronomy benchmark, strong API-served general-purpose
models establish a high correctness baseline, while specialized open-weight
models offer different properties related to local deployment, reproducibility,
controllability, and data governance. As reflected in
Tables~\ref{tab:usaaao-results} and~\ref{tab:compute-config}, these systems also
operate under different computational and serving environments. Model selection
for scientific workflows therefore involves both capability and deployment
requirements that are not represented by aggregate accuracy alone.

This comparison should not be interpreted as a definitive test of domain
adaptation itself. The astronomy-specialized models evaluated here are not
domain-adapted versions of the strongest contemporary general-purpose models,
and differences in base model capability, scale, training data, and adaptation
strategy remain confounding factors. Rather, the results raise a more targeted
question: whether adapting stronger contemporary open models can improve
domain-specific reasoning while retaining the access and deployment properties
that motivate specialization in the first place.

Our evaluation also demonstrates that the answer depends on how scientific
capability is measured. Reference-based metrics show varying agreement with
judge-based correctness, evaluator choice affects some model scores, and
performance varies across benchmark composition, question length, and modality.
These differences argue against reducing open-ended scientific evaluation to a
single metric or ranking. Instead, complementary evaluation views can help
identify where models succeed, where they fail, and whether observed differences
are robust to the characteristics of the scientific task.

More broadly, this study illustrates a growing challenge for AI for Science:
model capabilities, architectures, and deployment options are evolving faster
than any fixed benchmark can fully characterize them. Domain-specific evaluation
can therefore serve not only to rank models, but to identify which capabilities,
evaluation criteria, and deployment properties matter for a particular scientific
workflow. Repeating such characterization across scientific domains can help
establish where general-purpose models are sufficient, where specialization or
additional scaffolding provides measurable value, and which evaluation signals
are appropriate for scientific use.

Astronomy provides one controlled setting for this analysis, but the broader
methodology extends naturally to domains involving specialized knowledge,
quantitative reasoning, multimodal evidence, and domain-specific computational
constraints. As AI becomes more deeply integrated into scientific workflows,
such evaluations can provide empirical guidance for choosing among general
models, domain adaptation, retrieval, tool use, and other forms of scientific
AI augmentation rather than assuming that any one approach is universally
preferred.

\section{Limitations}

This study focuses on the free-response benchmark; the multiple-choice track is
reserved for future evaluation. Text-only and full-setting results are based on
different subsets and therefore should not be interpreted as direct comparisons.
Similarly, comparisons between API-served and open-weight models characterize
task performance rather than the full set of deployment tradeoffs.

Because the source materials are publicly available, potential training-data
contamination cannot be ruled out, particularly for broadly web-trained
models~\cite{deng2024contamination}. Variation across exam years provides a
coarse temporal diagnostic but cannot establish whether individual problems
were encountered during training or distinguish contamination from changes in
question content and difficulty.

The LLM-as-judge evaluation is also sensitive to judge selection and rubric
design. Our second-judge analysis provides an initial robustness check but does
not replace expert adjudication or a broader multi-judge evaluation. Likewise,
the targeted error analysis covers only a subset of responses and should not be
interpreted as an estimate of failure-mode prevalence across the full benchmark.

Finally, local and API-served models differ in serving infrastructure, context
handling, generation policies, and available hardware information. Provider-side
resource utilization is unavailable, and this study does not provide a complete
comparison of inference cost, GPU-hours, or energy consumption across serving
environments.

\section{Future Work}

Future work will extend the benchmark across task formats, model adaptation
strategies, and measures of scientific reliability. Incorporating the reserved
multiple-choice track will enable comparison between constrained-answer accuracy
and open-ended reasoning within the same astronomy testbed. We also plan to
expand the multimodal evaluation with additional vision-capable models to better
isolate visual reasoning from differences in problem content and difficulty.

A key direction is evaluating astronomy adaptation using stronger contemporary
open-weight base models. This will help separate the effects of domain
specialization from differences in model scale and underlying capability. Beyond
aggregate correctness, future evaluations will examine targeted scientific
properties such as topic robustness, physical consistency, multi-step reasoning,
and uncertainty calibration, including whether model confidence reliably tracks
correctness across problem difficulty.

We further plan to evaluate retrieval- and agent-based approaches that combine
language-model reasoning with astronomy resources, task decomposition, scientific
tools, and numerical or unit verification. These experiments can help determine
when domain adaptation, retrieval, tool use, or agentic scaffolding provides
measurable benefits over stronger general-purpose baselines, and whether those
benefits justify additional training, computational, and deployment costs.

\section{Conclusion}

We presented a benchmark for open-ended astronomy question answering using
Olympiad-style problems from 2017--2026 and used it to characterize
general-purpose and astronomy-specialized language models under a common
scientific reasoning task. The evaluation considers aggregate correctness,
modality, metric agreement, judge sensitivity, and robustness across benchmark
characteristics, providing a broader view of model behavior than a single
leaderboard.

Strong general-purpose models establish a high correctness baseline in this
testbed, while the value of domain specialization remains dependent on model
capability, task requirements, and deployment constraints. Our analyses also
show that reference-based similarity does not consistently reflect judged
scientific correctness and that small performance differences can be sensitive
to evaluator choice and benchmark composition.

More broadly, these results motivate domain-specific evaluation as a means of
identifying which models, capabilities, and evaluation criteria are appropriate
for scientific workflows. As AI systems continue to evolve, such characterization
can help determine where general-purpose models are sufficient and where domain
adaptation or additional scientific scaffolding provides measurable value.

\section{LLM Usage Disclosure}

Large language models assisted with code generation, data-format validation, figure generation, and drafting. The authors reviewed conversion scripts, JSONL records, result tables, figures, and manuscript text. Benchmark predictions were generated by the evaluated systems and scored with the fixed pipeline described above. The authors remain responsible for the final content, analysis, and conclusions.

\section{Artifact Description}

Code and analysis workflows are maintained at
\url{https://github.com/vanessalama09/astrobench}. The artifact supports
benchmark construction, evaluation, and reproduction of tables and figures from
per-example results. Model regeneration requires appropriate checkpoints,
compute resources, or API access. The structured USAAAO dataset will be released
following documentation and licensing review.

\section*{Acknowledgment}

This research used resources of the Oak Ridge Leadership Computing Facility (OLCF), which is a DOE Office of Science User Facility at the Oak Ridge National Laboratory supported by the U.S. Department of Energy under Contract No. DE-AC05-00OR22725. Tirthankar Ghosal was supported by the U.S. Department of Energy, Advanced Scientific Computing Research, through the SciDAC-RAPIDS3 institute.

\section*{Notice of Copyright}
This manuscript has been authored by UT-Battelle, LLC, under contract DE-AC05-00OR22725 with the US Department of Energy (DOE). The US government retains and the publisher, by accepting the article for publication, acknowledges that the US government retains a nonexclusive, paid-up, irrevocable, worldwide license to publish or reproduce the published form of this manuscript, or allow others to do so, for US government purposes. DOE will provide public access to these results of federally sponsored research in accordance with the DOE Public Access Plan ( https://www.energy.gov/doe-public-access-plan ).

\clearpage
\bibliographystyle{IEEEtran}
\bibliography{references}

\clearpage
\section*{Appendix: Additional Benchmark Details}

\subsection*{Specialization Deltas}
Specialization should be compared against nearby general baselines, not only against frontier systems. Table~\ref{tab:specialization-deltas} reports text-only judge-score differences between each astronomy-specialized model and the closest evaluated general model at a similar scale.

\begin{table}[!t]
  \caption{Specialization deltas on the shared text-only subset. $\Delta_{\mathrm{spec}}$ is specialized model score minus the nearest evaluated general baseline score.}
  \label{tab:specialization-deltas}
  \centering
  \small
  \setlength{\tabcolsep}{3pt}
  \begin{tabular}{lcc}
    \toprule
    Specialized model & Baseline & $\Delta_{\mathrm{spec}}$ \\
    \midrule
    AstroSage 70B & Llama 3.1 70B & +0.0735 \\
    AstroSage 8B & Llama 3.1 8B & +0.0833 \\
    AstroLLaMA 3 8B AIC & Llama 3 8B Inst. & -0.0368 \\
    \bottomrule
  \end{tabular}
\end{table}

\subsection*{Prompt Template and Judge Rubric}
The answer-generation prompt was:
\begin{quote}
\small
Answer the astronomy/astrophysics question directly. Provide a concise final answer with essential reasoning only. If the question is quantitative, show the key calculation. Question: \texttt{\{question\}}
\end{quote}

The judge received the question, reference answer, and model answer, and was instructed to return strict JSON with keys \texttt{verdict}, \texttt{score}, and \texttt{rationale}. Verdicts were restricted to \texttt{correct}, \texttt{partial}, or \texttt{incorrect}; scores were restricted to 1.0, 0.5, or 0.0; and rationales were required to be brief and factual.

\subsection*{Example Benchmark Questions}

\begin{figure}[!t]
  \centering
  \textbf{(a) Text-only example}\\[0.4ex]
  \includegraphics[width=\linewidth]{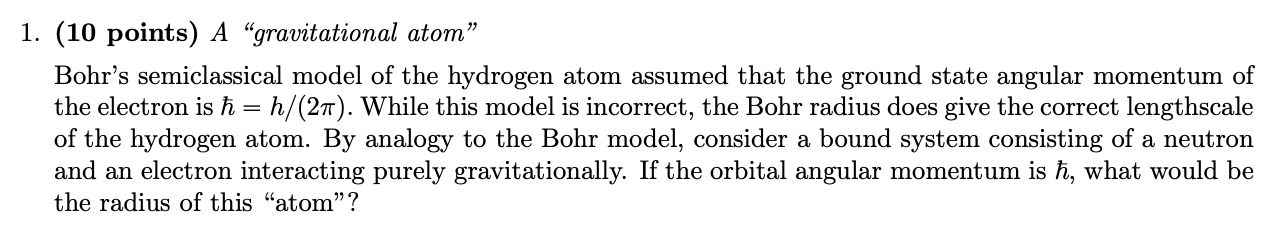}\\[1.0ex]
  \textbf{(b) Image-linked example}\\[0.4ex]
  \includegraphics[width=\linewidth]{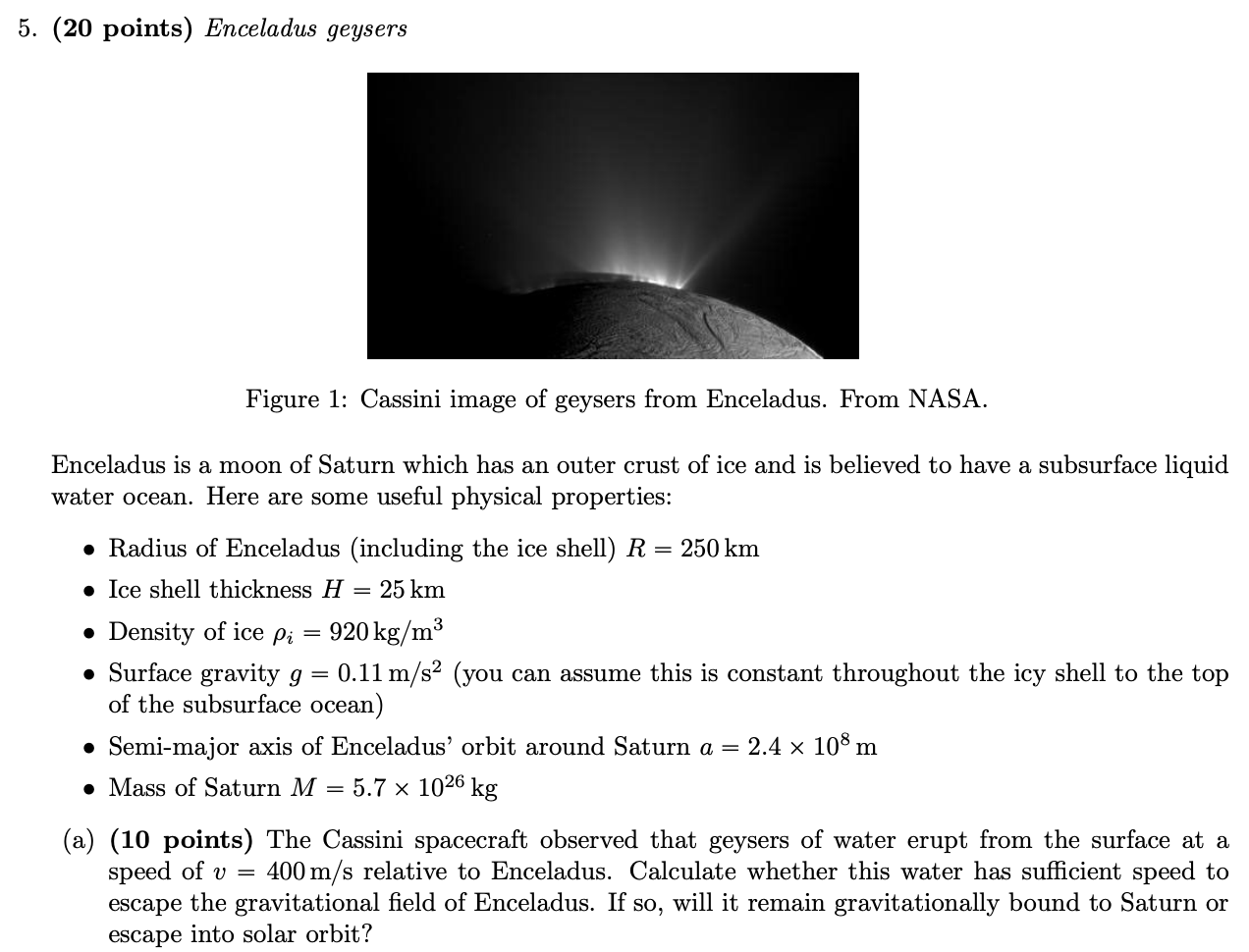}
  \caption{Representative 2026 free-response examples illustrating text-only and image-linked benchmark records.}
  \label{fig:example-questions}
\end{figure}

Fig.~\ref{fig:example-questions} shows representative 2026 text-only and image-linked examples. Both follow the JSONL record structure described in the methodology, with question text, reference answer, year, identifier, length bucket, and optional image paths.

\end{document}